\documentclass[superscriptaddress,twocolumn,notitlepage,showkeys,showpacs,longbibliography]{revtex4-1}
\usepackage[utf8]{inputenc}
\usepackage{stmaryrd}
\usepackage{amsmath}
\usepackage{amssymb}
\usepackage{graphicx}
\usepackage{dcolumn}
\usepackage{bm}
\usepackage{multirow}
\usepackage[colorlinks,linkcolor=blue,citecolor=blue,urlcolor=blue]{hyperref}
\usepackage{color}
\usepackage{ulem}
\usepackage{comment}
\usepackage{booktabs}
\usepackage{appendix}

\begin{document}
\title{Disordered charge density waves in the kagome metal FeGe}
\author{Hengxin Tan}
\author{Binghai Yan}
\affiliation{Department of Condensed Matter Physics, Weizmann Institute of Science, Rehovot 7610001, Israel}

\begin{abstract}
The discovery of a charge density wave (CDW) in the antiferromagnetic kagome metal FeGe has prompted interest in the interplay between kagome physics, CDW, and magnetism. However, a crucial aspect for understanding these emergent phenomena—the precise CDW structure—remains ambiguous. Recent studies have assumed uniformly distributed Ge dimers, but this assumption is problematic. The predicted band structure based on this model exhibits an abrupt disappearance of a Ge-$p$ band in the Fermi surface, contradicting experimental observations from angle-resolved photoemission spectroscopy (ARPES). In this study, we propose that a CDW phase with disordered Ge dimers can reconcile theoretical predictions with ARPES results. This model reproduces the observed CDW gaps while preserving the Ge-$p$ band. Depending on experimental conditions, Ge dimers can be randomly distributed or exhibit phase separation from pristine regions. Our findings reveal the crucial role of Ge dimer disorder in the FeGe CDW and suggest potential implications of this disorder for other properties, such as magnetism and transport, in this system.
\end{abstract}

\maketitle

\textit{Introduction. }
The recent discovery of a charge density wave (CDW) in the magnetic kagome material FeGe has opened new avenues for exploring the interplay between magnetism and electronic instabilities in this distinctive lattice type \cite{teng2022discovery}.
As temperature decreases, FeGe undergoes several phase transitions: an A-type antiferromagnetic (AFM) transition at approximately 410 K, a 2$\times$2$\times$2 CDW transition around 100 K, and a spin-canting phase at about 60 K.
The CDW phase plays a crucial role in elucidating the interactions between multiple exotic phases in FeGe and related kagome metals. Despite extensive research on FeGe \cite{shao2023intertwining,zhou2023magnetic,wu2023electron,miao2023signature,wang2023enhanced,zhang2023electronic,yin2022discovery,setty2022electron,teng2023magnetism,zhao2023photoemission,wenzel2024intriguing,ma2023theory,teng2024spinchargelattice,yi2024polarized,chen2024competing}, the exact configuration of the CDW phase remains elusive, hindering the complete understanding of the underlying physics.

Earlier first-principles calculations proposed a 2$\times$2$\times2$ CDW structure characterized by the Ge-honeycomb layer Kekule distortion and Ge-Ge dimers along the out-of-plane direction \cite{shao2023intertwining}. Subsequent research identified that Ge dimers play an essential role in the structural distortion \cite{miao2023signature,wang2023enhanced}. This distortion enhances spin polarization \cite{shao2023intertwining,wang2023enhanced}, which qualitatively aligns with experimental observations \cite{teng2022discovery}. Additionally, the predicted 1/4 Ge dimer ratio in kagome layers is consistent with experimental findings \cite{chen2024discovery}. Therefore, this structure is a strong candidate for the CDW structure in FeGe. 
However, it shows a significant discrepancy in the electronic structure compared to angle-resolved photoemission spectroscopy (ARPES) experiments. 
In calculations, this CDW structure provides a profound upshift of the Ge $p_z$ band by about 0.5 eV at the Brillouin zone center and consequently removes its Fermi surface compared to the pristine phase \cite{shao2023intertwining,miao2023signature,wang2023enhanced,zhang2023electronic,zhao2023photoemission}. In contrast, this band remains nearly unchanged by remaining a large Fermi surface during the CDW transition in ARPES \cite{zhao2023photoemission,teng2023magnetism}.
This key difference in the electronic structure raises doubts about the presumed CDW structure.

On the other hand, while initial experiments observed a short-range CDW \cite{teng2022discovery}, recent experiments found that the coherence length of the CDW is tunable by sample annealing, allowing for the achievement of a long-range CDW \cite{chen2024discovery,oh2024tunability}. Sample post-processing, such as annealing, typically addresses structural disorders, defects, and domains. Indeed, recent experiments have identified significant Ge disorder in FeGe \cite{wu2023annealingtunable,shi2023disordered}. This motivates us to evaluate the effect of the atomic disorder on the electronic structure of FeGe.

In this work, we find that non-uniformly distributed Ge dimers reproduce key features of ARPES, i.e., showing CDW gaps near the van Hove singularity and preserving the Ge $p_z$ band at the Fermi surface.
In contrast to the ordered dimer configuration, we propose two disorder scenarios. The first involves randomly distributed dimers that disrupt phase coherence and blur band dispersion while maintaining the original Fermi surface. The second scenario posits a phase separation between pristine and dimer regions, where the pristine regions contribute to the preservation of the Ge $p_z$ bands. Both scenarios are hinted in recent experimental evidence, suggesting that the CDW structure is dependent on sample conditions. Our results indicate that Ge dimer disorders may commonly exist and affect the electronic and magnetic properties of FeGe and similar kagome materials.

\begin{figure*}[tbp]
\includegraphics[width=0.8\linewidth]{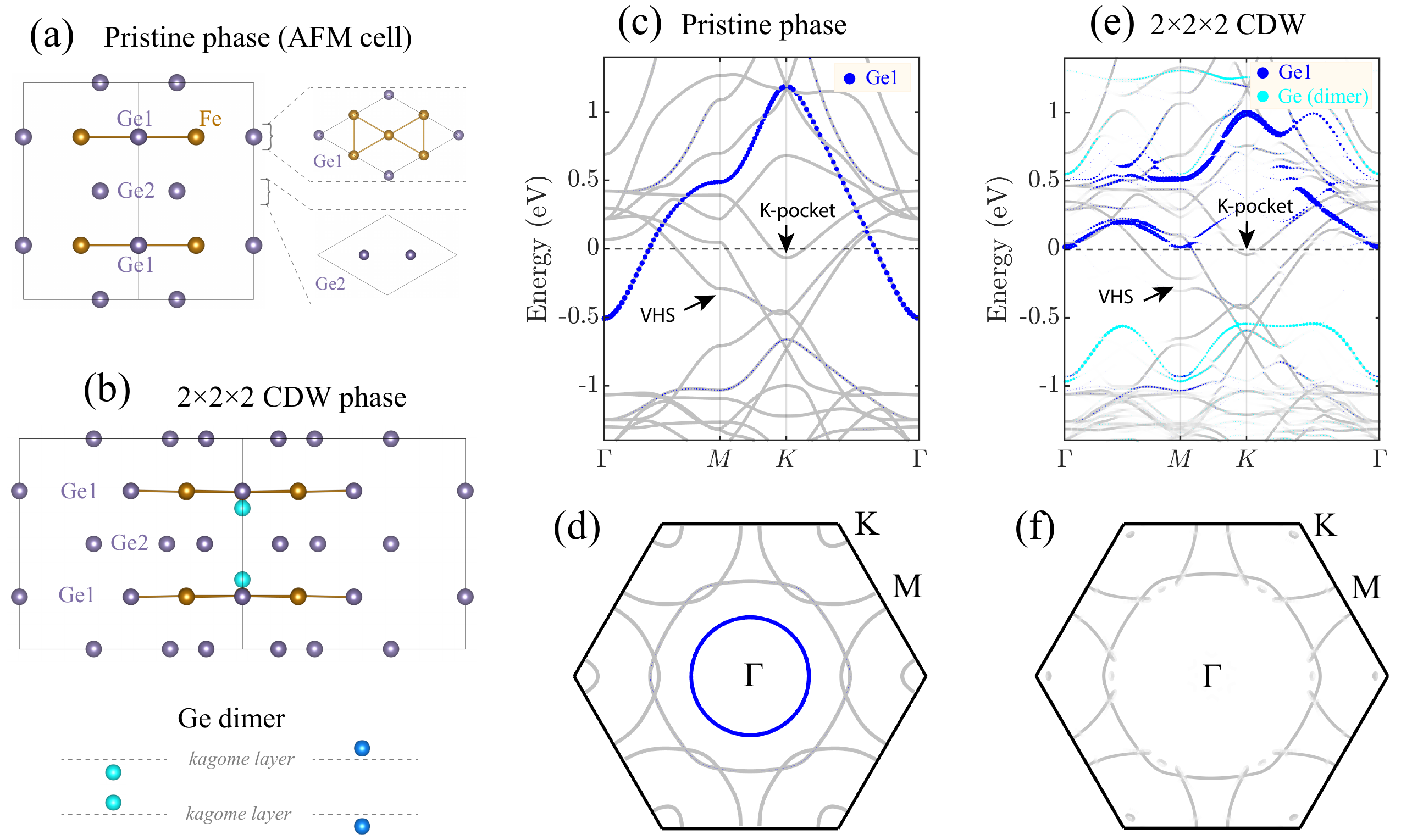}
\caption{\label{Fig1}
(a) The AFM cell (1$\times$1$\times$2) of the pristine phase FeGe in the side view. The kagome (Fe-Ge1) and honeycomb (Ge2) sublayers are displayed.
(b) The candidate CDW structure (2$\times$2$\times$2) with a dimer (cyan) formed by a pair of Ge1 atoms. Depending on the displacement directions of Ge1 atoms, two dimer configurations distinguished by colors cyan and navy are shown in the bottom row.
(c) and (d) Band structure and Fermi surface on $k_z$=0 plane of the pristine AFM FeGe, respectively. Blue dots and curves show the Ge1 $p_z$ projection.
(e) and (f) Unfolded band structure and Fermi surface of the 2$\times$2$\times$2 structure shown in (b).
Cyan dots in (e) show the band weight of the Ge dimer; blue dots show the remaining Ge1 atoms. The dot size is proportional to the band weight.
The Fe-$d_{xz}/d_{yz}$ contributed $K$-point electron pocket and most relevant van Hove singularity (VHS) at $M$ are highlighted in (c) and (e).
}
\end{figure*}

\textit{Ordered Ge dimers. }
The crystal structure of FeGe is shown in Fig. \ref{Fig1}(a).
It comprises one Fe kagome layer with additional Ge1 atoms positioned at the hexagon centers and one Ge2 honeycomb layer.
Since FeGe has the A-type AFM configuration (interlayer AFM and intralayer ferromagnetic), we utilize the 1$\times$1$\times$2 supercell for the pristine phase to accommodate the magnetic order, and all band structures are plotted in its Brillouin zone.
If a pair of Ge1 atoms from two neighboring kagome layers in a 2$\times$2$\times$2 supercell move toward each other by $\sim$0.64 \AA~(from density-functional theory, DFT, calculations) to form a dimer (movements of other atoms are smaller than 0.1 \AA), the candidate CDW structure arises in Fig. \ref{Fig1}(b).
In experiments \cite{wu2023annealingtunable,shi2023disordered}, the Ge disorder mainly happens on Ge1 sites. Thus, we focus on the Ge1 site in the following.

The band structure of the pristine phase is shown in Fig. \ref{Fig1}(c). We consider only the $k_z$=0 plane, corresponding to overlapping $k_z = 0$ and $\pi$ planes of ARPES results because we use the $1\times1\times2$ AFM structure in calculations. 
While most bands originate from Fe $d$, the Ge1 $p_z$ contributes a dispersive band across the Fermi energy, forming a circular Fermi surface in Fig. \ref{Fig1}(d).
When one pair of Ge1 atoms form dimers in the 2$\times$2$\times$2 structure, the $p_z$ band from the non-dimerized Ge1 atoms at $\Gamma$ point moves upward by about 0.5 eV right above the Fermi level as shown in Fig. \ref{Fig1}(e); simultaneously, the $p_z$ band from Ge dimers also splits, going far away from the Fermi level, consistent with previous calculations \cite{zhang2023electronic}. Consequently, the circular $p_z$ Fermi surface is absent in the 2$\times$2$\times$2 Ge dimerized structure, as depicted in Fig. \ref{Fig1}(f).

As mentioned above, such a significant band modification contradicts ARPES experiments. Zhao $et$ $al.$ \cite{zhao2023photoemission} observed only $\sim$25 meV upward shift of the Ge1 $p_z$ band at $\Gamma$, and Teng $et$ $al.$ \cite{teng2023magnetism} revealed no noticeable Fermi surface change across the CDW transition.
Ge dimers are most uniformly distributed in a 2$\times$2 in-plane supercell at the consensus 1/4 dimer ratio.
In this model, ordered dimers form a periodic lattice, substantially modifying the original band structure and leading to a significant upward shift of the $p_z$ band. Based on this interpretation, we propose two scenarios that could explain the minor band modifications observed in experiments:
(i) The existence of dimer-free regions that weakly interact with Ge dimers, resulting in phase separation between pristine and dimer regions.
(ii) Randomly distributed dimers that reduce the phase coherence of the original lattice.
In both scenarios, the $p_z$ band dispersion would remain relatively stable during the CDW transition. These two scenarios represent extreme disorder conditions when keeping the fixed dimer ratio of 1/4.

\begin{figure}[tbp]
\includegraphics[width=0.95\linewidth]{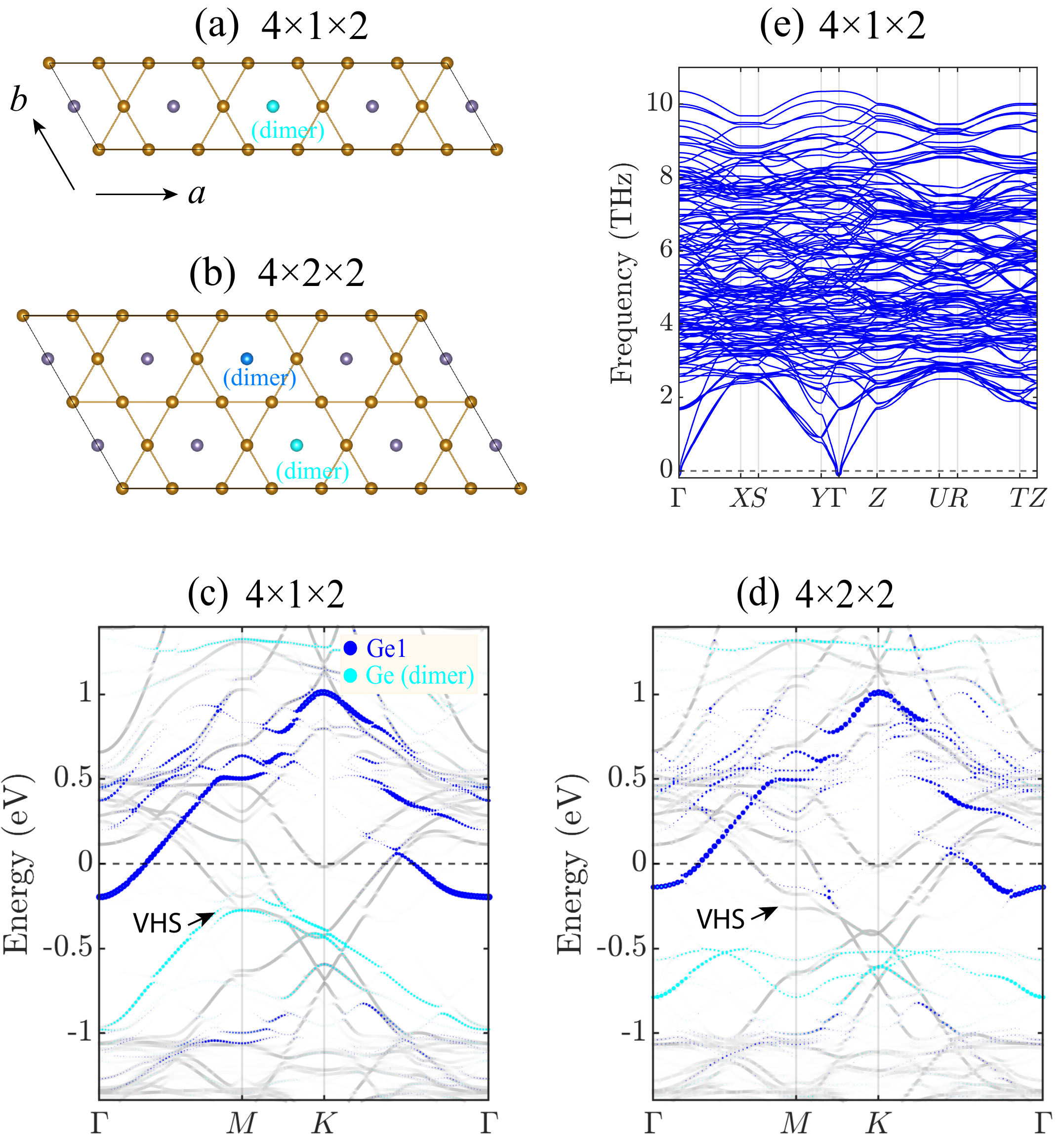}
\caption{\label{Fig2}
(a)-(d) Crystal structures of two non-uniformly distributed Ge dimer models in two supercells 4$\times$1$\times$2 and 4$\times$2$\times$2 in the top view. For simplicity, the Ge2 honeycomb layers are not shown. Notice that a pair of Ge1 can move toward (cyan color) or backward (dark blue color) each other along $c$ to form dimers between different kagome layers [see Fig. \ref{Fig1}(b)].
(c)-(d) Unfolded energy dispersions (grey) of these two structures. Ge $p_z$ band weights are overlaid, where Ge dimer refers to both types.
(e) Phonon dispersion of the 4$\times$1$\times$2 structure in (a). High-symmetry $q$ points are standard points in a simple orthorhombic system.
}
\end{figure}

\textit{Phase separation.}
We will first showcase the phase separation scenario by employing specific dimer configurations. Next, by extracting an effective model Hamiltonian for the Ge band, we will demonstrate the random disorder scenario using large supercells.
Figure \ref{Fig2}(a)\&(b) shows two strip Ge dimer models in 4$\times$1$\times$2 and 4$\times$2$\times$2 supercells with a fixed dimer ratio of 1/4.
Notice that Ge1 atoms from two neighboring kagome layers can move toward or backward each other along the out-of-plane direction (lattice vector $c$) to form dimers between different kagome layers [see Fig. \ref{Fig1}(b)]. 
These models have the most non-uniform but ordered dimer distribution in the corresponding supercells$-$dimers are most densely packed along the $b$ axis and sparsely packed along the $a$ axis.
Corresponding DFT band structures in Fig. \ref{Fig2}(c)\&(d) show a much less upward shift of the non-dimerized Ge1 $p_z$ band at $\Gamma$, as compared to the 2$\times$2$\times$2 structure in Fig. \ref{Fig1}(e).
These results indicates the phase separation can maintain the $p_z$ band because of existence a large pristine area.
We note that the adjacent dimer Ge1 atoms moving toward or backward each other do not generate essential differences in the band structure, though the two configurations have a noticeable energy difference (see Table \ref{Table1}).
The dynamic stability of the 4$\times$1$\times$2 supercell is proved by the absence of imaginary phonon models in its phonon dispersion, depicted in Fig. \ref{Fig2}(e).
The average local magnetic moment of Fe atoms in both structures is enhanced by 0.08 $\mu_B$ compared to the pristine phase, in line with the experiment \cite{teng2022discovery}.

In addition to maintaining the Ge $p_z$ band, our calculations also qualitatively reproduced other key features of ARPES experiments.
(i) The CDW gap. As shown in Fig. \ref{Fig2}(c)\&(d), the two strip dimer models open a gap for the van Hove singularity (VHS) at $M$ below the Fermi level, which agrees with recent experiments \cite{teng2022discovery,teng2023magnetism}. We mention that the 2$\times$2$\times$2 superstructure also opens a gap for the VHS in Fig. \ref{Fig1}(e).
(ii) Energy shift of the $K$-point pocket. The experiment \cite{zhao2023photoemission} also observed a slight elevation ($\sim$28 meV) of the $K$-point electron pocket [highlighted in Fig. \ref{Fig1}(c)\&(e)] across the CDW transition. The upward shift of this pocket, composed of Fe $d_{xz}/d_{yz}$ orbitals, is captured by both the 2$\times$2$\times$2 candidate CDW structure and our dimer models.

\begin{figure*}[tbp]
\includegraphics[width=0.95\linewidth]{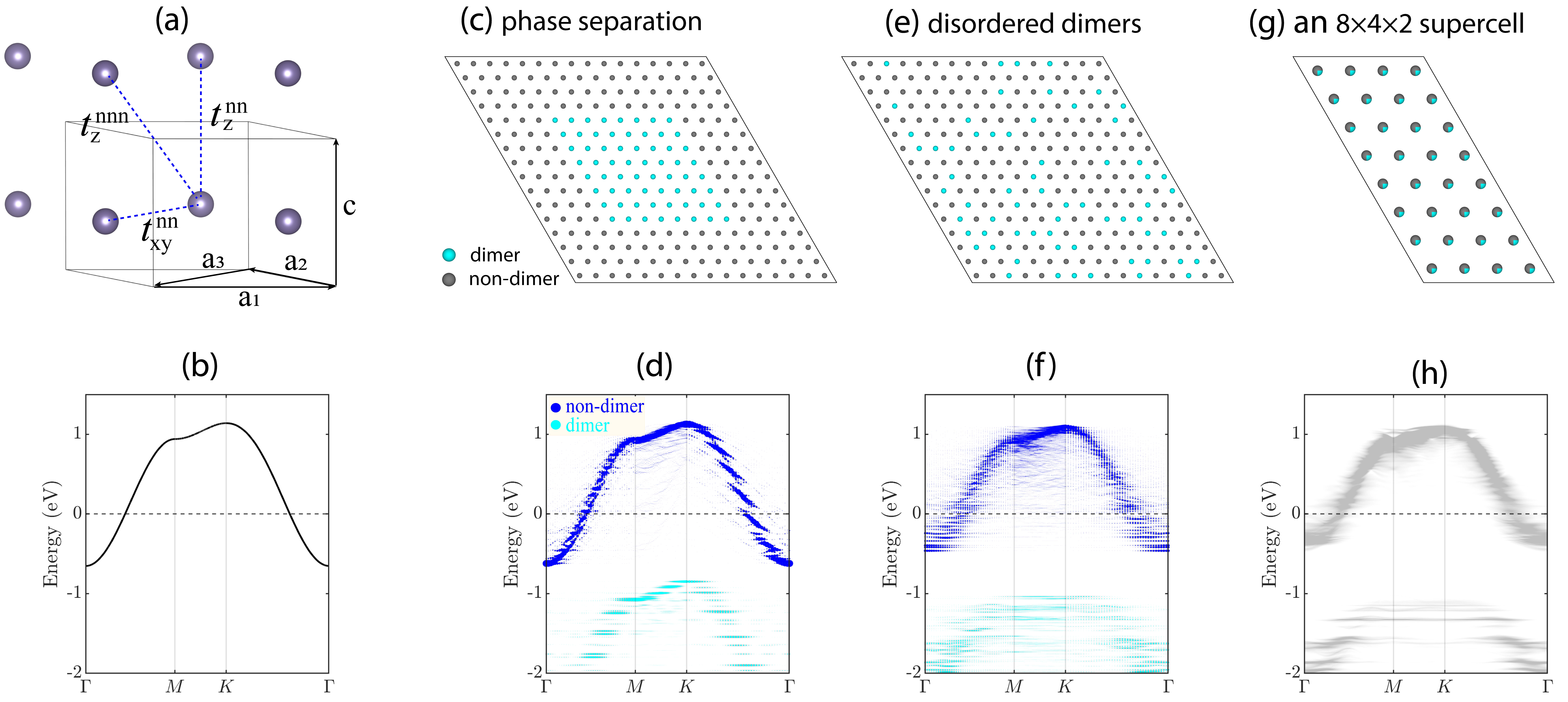}
\caption{\label{Fig3} 
Model calculations of disordered dimers. (a) The 3D triangular lattice simulated in the tight-binding model with parameters labeled. (b) The band structure of the pristine phase from the model with parameters \{$t_{xy}^{nn}, t_{z}^{nn}, t_{z}^{nnn}, U$\} = \{$-0.05, 0.479, 0.075, -1.5$\} eV, which is similar to the blue Ge1 band in Fig. \ref{Fig1}(c).
(c) A representative phase separation between the pristine and dimer regions caused by dimer disorder in a 16$\times$16$\times$2 supercell.
(d) The unfolded band structure of the phase separation configuration in (c).
(e) A fully disordered dimer distribution obtained from the special quasirandom structure method \cite{Zunger1990special} in the same supercell.
(f) The corresponding band structure of the disordered dimer distribution in (e).
(g) An 8$\times$4$\times$2 supercell of the model under the 1/4 dimer ratio, with the average band structure of all possible dimer configurations ($\sim10^7$) in (h).
Structures (c), (e), and (g) are shown in the top view (dimer ratio fixed at 1/4). $\lambda=6$ is adopted in band structure calculations in (d), (f), and (h).
}
\end{figure*}

\textit{Disordered dimers.}
According to our DFT calculations for various dimer models shown in the Supplemental Material (SM) \cite{SM}, the major changes in the band structure caused by Ge dimers are manifested by the Ge $p_z$ band, which is the focus of this section. We will first develop a one-band tight-binding model for the 3D triangular lattice of Ge1 [Fig. \ref{Fig3}(a)] to fit the Ge $p_z$ band and then evaluate the disorder effect based on the model. The $c$/$a$ ratio of the model is fixed at the realistic value 0.82 of FeGe ($a$ and $c$ are the in-plane and out-of-plane lattice constants). By taking $p_z$ as the basis and including hoppings to the third nearest neighbors, the one-band tight-binding Hamiltonian reads,
\begin{equation}
\begin{split}
\begin{aligned}
    H = & \sum t_{xy}^{nn} \hat c^{\dag}(\textbf{R}) \hat c(\textbf{R}+\textbf{r}_{xy}) + \sum t_{z}^{nn} \hat c^{\dag}(\textbf{R}) \hat c(\textbf{R}+\textbf{r}_{z}) \\
    & + \sum t_{z}^{nnn} \hat c^{\dag}(\textbf{R}) \hat c(\textbf{R}+\textbf{r}_{xy}+\textbf{r}_z) + U \sum \hat c^{\dag}(\textbf{R}) \hat c(\textbf{R}),
\nonumber
\end{aligned}
\end{split}
\end{equation}
where $\hat c^{\dag}(\textbf{R})$ and $\hat c(\textbf{R})$ are the creation and annihilation operators respectively.
The first term corresponds to the in-plane nearest-neighboring (nn) hopping ($t_{xy}^{nn}$),
the second term stands for the out-of-plane nn hopping ($t_{z}^{nn}$), and the third term represents the out-of-plane next nn (nnn) hopping ($t_{z}^{nnn}$). 
$U$ is the on-site potential. $\textbf{r}_{xy}$ $\in \{\pm \textbf{a}_i$\} ($i=1,2,3$) and $\textbf{r}_z$ $\in \{\pm \textbf{c}$\}, where $\textbf{a}_i$ and $\textbf{c}$ are the in-plane and out-of-plane lattice vectors, respectively. $\textbf{a}_1=(1,0,0)$; $\textbf{a}_2=(-1/2,\sqrt3/2,0)$; $\textbf{a}_3=(-1/2,-\sqrt3/2,0)$; $\textbf{c}=(0,0,0.82)$.
The analytic expression of the Bloch Hamiltonian in reciprocal space reads,
\begin{equation}
\begin{split}
\begin{aligned}
    H(\textbf{k}) = & 2\sum_{i} \textrm{cos}(\textbf{k} \cdot \textbf{a}_i) [t_{xy}^{nn} + 2 t_{z}^{nnn} \textrm{cos}(\textbf{k} \cdot \textbf{c})] \\
    & + 2 t_{z}^{nn} \textrm{cos}(\textbf{k} \cdot \textbf{c}) + U.
\nonumber
\end{aligned}
\end{split}
\end{equation}
By fitting the Ge1 $p_z$ band of pristine FeGe, we obtain an optimal set of model parameters, \{$t_{xy}^{nn}, t_{z}^{nn}, t_{z}^{nnn}, U$\} = \{$-0.05, 0.479, 0.075, -1.5$\} eV \cite{note1}. Fig. \ref{Fig3}(b) shows the corresponding energy dispersion.

Now, we extend the toy model to a target supercell and manually create dimers via shifting 1/4 lattice sites by $\delta \sim0.15 \textbf{c}$ (0.15 is the relative dimer movement in FeGe to its primitive out-of-plane lattice constant). Simultaneously, the hopping integral between any two sites is modified exponentially,
\[ t = t_0 e^{\lambda (1-d/d_0)}. \]
Here, $t_0$ is the initial hopping integral; $\lambda$ is a scaling factor; $d_0$ and $d$ are the distances between the two sites before and after dimerization.
$d$ may have a complicated dependence on $\delta$. $d=d_0$ if both sites were irrelevant to dimers. After obtaining the revised Hamiltonian, we unfold the band structure to the Brillouin zone of the 1$\times$1$\times$2 supercell according to the method in Ref. \onlinecite{Deretzis2014role}.
Using DFT results in Fig. \ref{Fig1}(e) and Figs. \ref{Fig2}(c)\&(d) for benchmarks, we obtained an overall optimal value $\lambda \sim 6$ (corresponding band structures are shown in Fig. S3 \cite{SM}). This simple toy model thus enables the exploration of the Ge1 $p_z$ band with dimer disorder in large supercells.
Notice that a similar procedure is applied for the realistic wannier Hamiltonian of FeGe, which also captures the DFT band structures well (Fig. S2 in SM \cite{SM}).
However, due to the large number of wannier orbitals in the realistic model, it is not feasible to deal with dimer disorder in very large supercells.

We extend the above model to a 16$\times$16$\times$2 supercell to incorporate dimer disorders more realistically (more results are in Fig. S4 \cite{SM}). Two dimer configurations are considered, as shown in Fig. \ref{Fig3}(c) and (e): a representative phase separation between the pristine and dimer regions, and a fully disordered configuration with weak coherence effect obtained using the special quasirandom structure method \cite{Zunger1990special} implemented in the \textsc{atat} software \cite{Walle2002the}. Their respective band structures are shown in Fig. \ref{Fig3}(d) and (f), which are largely similar. The only notable difference is that the band structure of the disordered configuration in Fig. \ref{Fig3}(f) appears more blurred compared to Fig. \ref{Fig3}(d), a consequence of the weaker coherence of hopping parameters in the disordered model. Most importantly, the non-dimer-related branch in both band structures closely resembles that of the clean lattice without dimers, as shown in Fig. \ref{Fig3}(b), indicating negligible energy shift. Besides, we average band structures of all possible dimer configurations in an 8$\times$4$\times$2 supercell as shown in Fig. \ref{Fig3}(g), which shows similar results in Fig. \ref{Fig3}(h).
These findings support our hypothesis that both the phase separation and the fully random disorder maintain the Ge1 $p_z$ band in energy.

Notably, phase separation in FeGe was recently observed in a scanning tunneling microscope (STM) study \cite{chen2024discovery}. In comparison, the earlier STM experiment \cite{yin2022discovery} reported CDW without any phase separation. Additionally, disordered CDW nuclei have been reported \cite{chen2023charge}. These varied observations suggest that either phase separation or random disorder mechanisms may exist in experiments, depending on sample preparation conditions. It is worth noting that in the dimer models discussed above, the average rotational symmetry of the system (and the actual FeGe) is preserved due to the Ge dimer disorder, in line with the rotational symmetric Fermi surface observed by ARPES experiments \cite{teng2023magnetism}. We mention that a recent Raman study reported weak rotational symmetry breaking \cite{wu2024symmetry} before the CDW transition in FeGe.

Finally, we note that many disordered dimer configurations in FeGe should be dynamically stable.
The stability of the 4$\times$1$\times$2 structure is proved by the absence of imaginary phonon modes in Fig. \ref{Fig2}(e), even though the stronger dimer-dimer interaction in this structure increases its total energy by a considerable amount, as summarized in Table \ref{Table1}.
It is thus plausible to expect more dimer configurations with comparable or lower total energy that are stable and accessible in experiments. The numerous meta-stable states are also predicted in a recent DFT calculation \cite{wang2024encoding}.
The multiple meta-stable states on the energy surface of FeGe facilitate its structural disorder.
Besides, multiple CDW domains further complicate the disorder in FeGe \cite{shi2024charge}.

\begin{table}
\centering
\caption{\label{Table1} Total energy of different Ge dimer models from DFT calculations (in the unit of meV per AFM cell). The total energy of the pristine phase is taken as the energy zero.
Structures are shown in Fig. \ref{Fig2} and Fig. S2 in SM \cite{SM}.
}
\begin{ruledtabular}
\begin{tabular}{ccccccccccc}
    2$\times$2$\times$2 & 4$\times$1$\times$2 & 4$\times$2$\times$2 &4$\times$4$\times$2 & 4$\times$4$\times$2 & 8$\times$1$\times$2 & 8$\times$1$\times$2 \\
                        &                     &                     & (i)               & (ii) & (i) & (ii)\\
    \specialrule{0em}{1pt}{1pt}
    \hline
    \specialrule{0em}{1pt}{1pt}
    $-$1 & 20 & 4 & 28 & 21 & 38 & 19 \\
\end{tabular}
\end{ruledtabular}
\end{table}

\textit{Conclusion.} 
We have elucidated the impact of Ge dimer disorder on the electronic structure of FeGe through theoretical investigations. Our results demonstrate that the experimentally observed minor changes in the band structure across the CDW transition can be explained by the dimer disorder, through phase separation, random distribution, or a combination of both. This structural disorder coincides with the CDW controllability by sample post-processing. Our work establishes FeGe as an ideal platform for exploring the effects of structural disorder on properties such as magnetism and transport.

\textbf{Acknowledgement.}
We acknowledge helpful discussions with Ming Yi, Pengcheng Dai, Girsh Blumberg, and Xiaokun Teng. We thank Yufei Zhao for assistance with the calculations. B.Y. acknowledges the financial support by the Israel Science Foundation (ISF: 2932/21, 2974/23) and German Research Foundation (DFG, CRC-183, A02).


\begin{thebibliography}{35}
\bibitem [{not()}]{note1}%
  \BibitemOpen
  \href@noop {} {}\bibinfo {note} {The out-of-plane hopping parameters $t_z^{nn}$ and $t_z^{nnn}$ are larger than the in-plane ones $t_{xy}^{nn}$ due to the significantly greater overlap of $p_z$ orbitals along the out-of-plane direction. Additionally, note that these hopping parameters have opposite signs.}\BibitemShut {Stop}%
\end{thebibliography}
%

\end{document}


\title{Supplemental Material for ``Disordered charge density waves in the kagome metal FeGe''}
\author{Hengxin Tan}
\author{Binghai Yan}
\affiliation{Department of Condensed Matter Physics, Weizmann Institute of Science, Rehovot 7610001, Israel}

\maketitle

\section{methods}
DFT calculations are performed with the Vienna $ab$-$initio$ Simulation Package \cite{PRB54p11169,VASP}. The generalized gradient approximation parameterized by Perdew, Burke, and Ernzerhof (PBE) \cite{PRL77p3865} is employed to mimic the exchange-correlation interaction between electrons throughout.
The energy cutoff for the plane wave basis set is 350 eV.
All crystal structures are fully relaxed until the remaining forces are less than 1 meV/\AA. Different $k$-meshes are utilized for different structures so that the $k$-space resolution is $\sim$0.15 \AA$^{-1}$. DFT band unfolding is performed with the post-processing VASPKIT package \cite{vaspkit}. 
The realistic tight-binding Hamiltonian is obtained with the maximally localized Wannier functions method \cite{wannier90}, where Fe $d$ and Ge $p$ orbitals compromise the Wannier basis set.
Phonon calculations are performed with the \textsc{phonopy} software \cite{phonopy}.
Spin-orbital coupling is omitted in all calculations.

\section{More results}

\begin{figure}[tbp]
\includegraphics[width=0.99\linewidth]{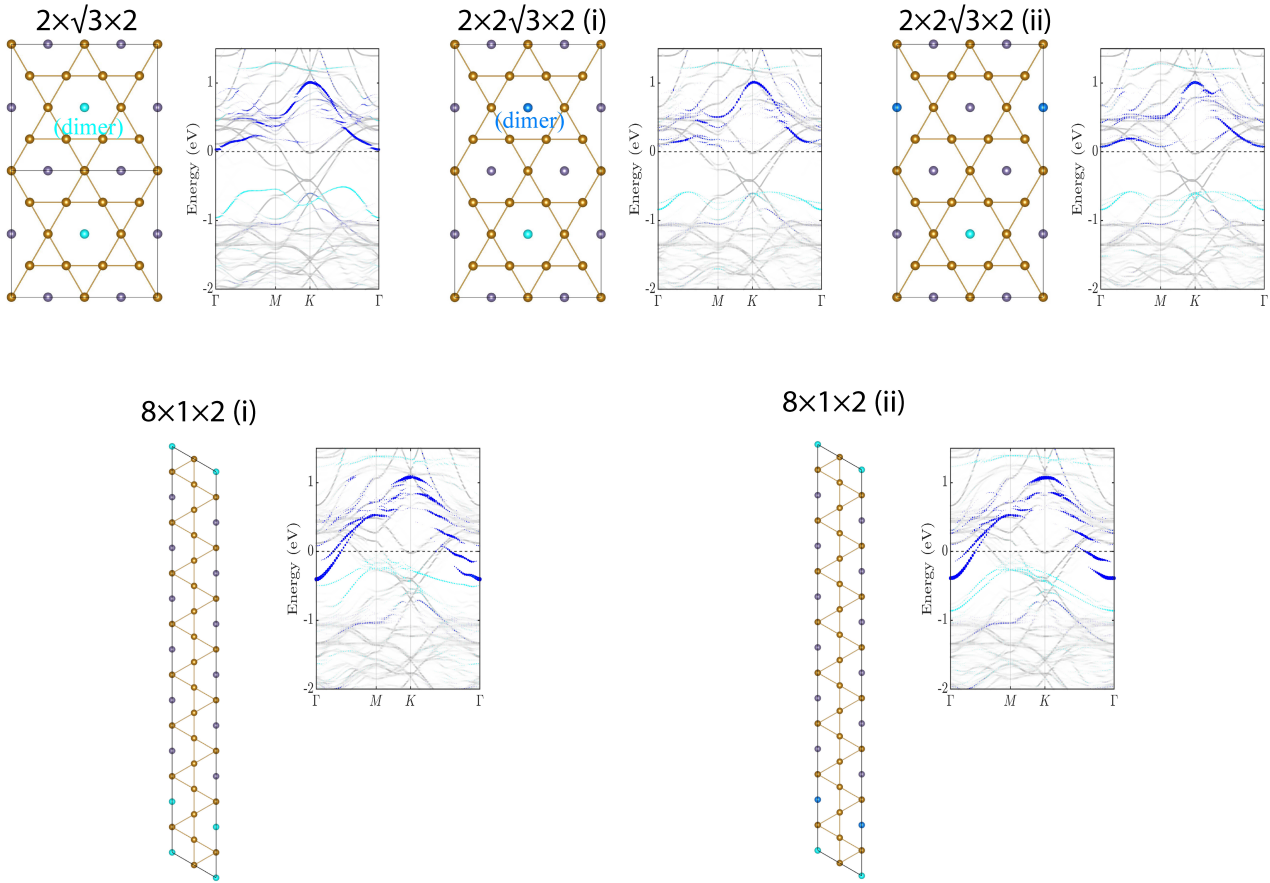}
\caption{\label{FigS1}
Dimer models and the corresponding unfolded DFT band structures.
The two different dimer types are defined in Fig. 1(b) in the main text.
For simplicity, only Ge1 and Fe atoms are shown in each structure.
}
\end{figure}

\begin{figure}[tbp]
\includegraphics[width=0.99\linewidth]{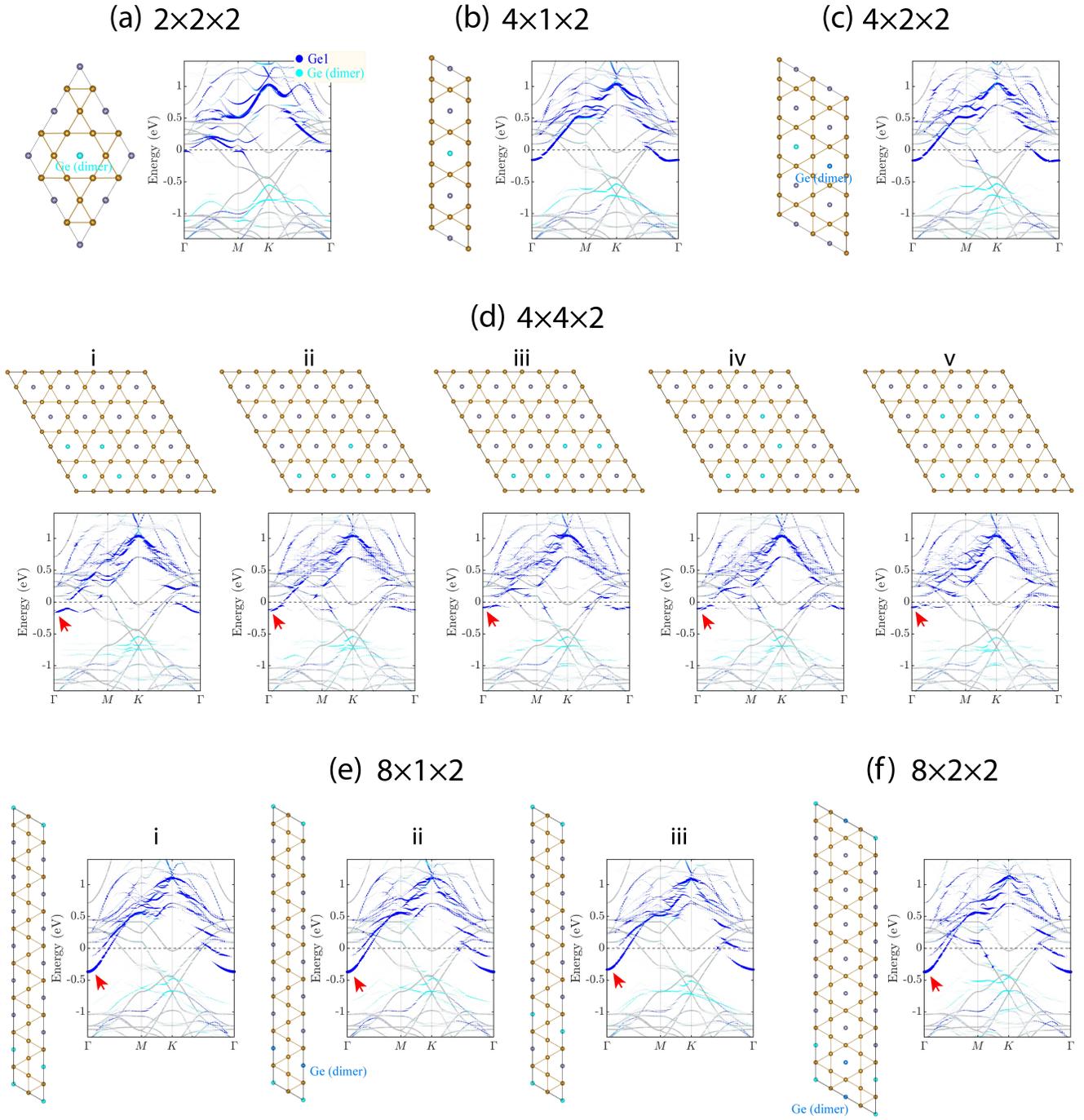}
\caption{\label{FigS2}
Different dimer models and corresponding band structures from the realistic Wannier Hamiltonian calculation.
The band structures are calculated in the following processes:
i) Extract the tight-binding Wannier Hamiltonian of the pristine phase of FeGe.
ii) Dimerized 1/4 Ge1 atoms in a supercell of the model and revised the hopping parameters of the corresponding supercell Hamiltonian according to $t = t_0 e^{\lambda (1-d/d_0)}$, with $\lambda$=6.
iii) Unfold the supercell band structure in the Brillouin zone of the pristine AFM cell.
For simplicity, only Ge1 and Fe atoms are shown in each structure.
}
\end{figure}

\begin{figure}[tbp]
\includegraphics[width=0.9\linewidth]{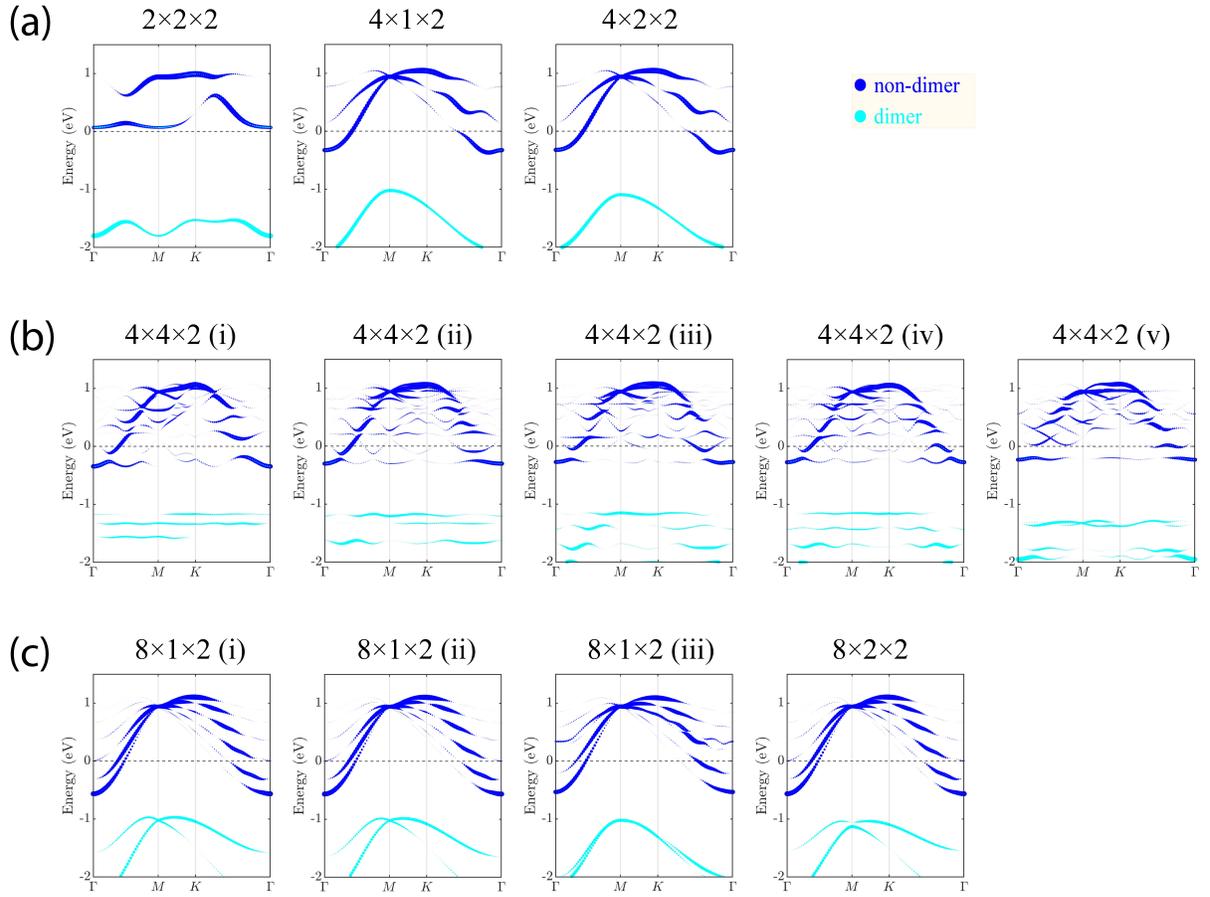}
\caption{\label{FigS3}
Band structures for the same dimer configurations in Fig. \ref{FigS2}, as obtained from the toy model calculations developed in the main text. $\lambda=6$ is employed.
}
\end{figure}

\begin{figure}[tbp]
\includegraphics[width=0.9\linewidth]{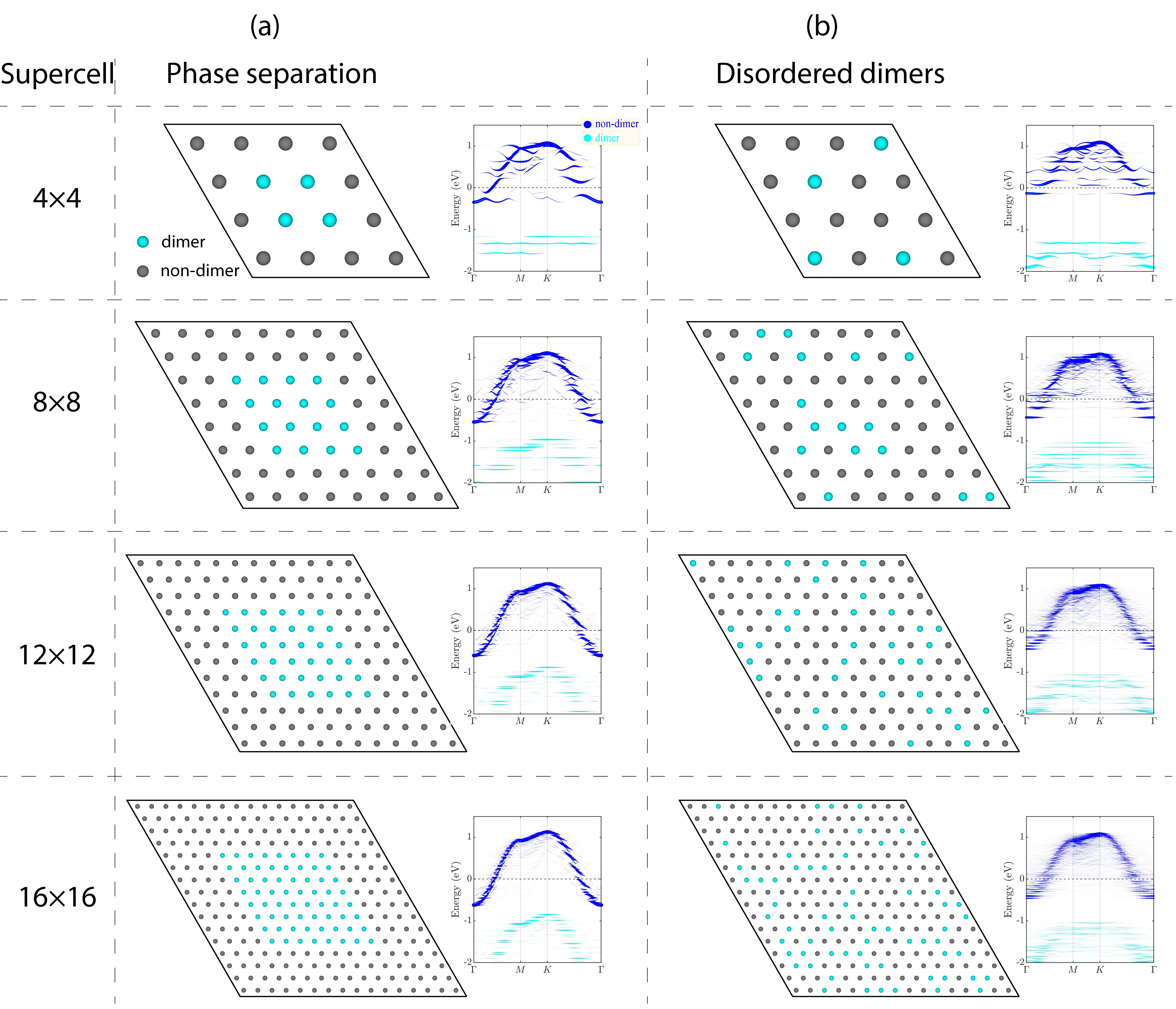}
\caption{\label{FigS4}
Phase separation (a) and fully disordered (b) dimer models in different supercells.
For phase separation, we consider a representative one composed of dimer clusters.
The disordered dimer models are obtained from the special quasirandom structure method \cite{Zunger1990special}.
Band structures are obtained from the toy model with $\lambda=6$. Results for the 16$\times$16 supercell are used as Fig. 3 in the main text.
Both dimer models maintain the Ge $p_z$ band below the Fermi energy at the $\Gamma$ point.
}
\end{figure}


\clearpage
%